\documentstyle[12pt,epsfig]{article}
\advance \topmargin by -\headheight
\advance \topmargin by -\headsep
\evensidemargin \oddsidemargin
\begin{document}
\pagestyle{plain}
\begin{titlepage}
\flushright{IHEP 2002-6}
\flushright{\today}
\vspace*{0.15cm}
\begin{center}
{\Large\bf
    Study of the $K^{-} \rightarrow \mu^{-} \bar{\nu} \pi^{0}  $ decay  
    }
\vspace*{0.15cm}

\vspace*{0.3cm}
{\bf  I.V.~Ajinenko, S.A.~Akimenko,  
G.I.~Britvich, I.G.~Britvich, K.V.~Datsko,  A.P.~Filin, 
A.V.~Inyakin,  A.S.~Konstantinov, V.F.~Konstantinov, 
I.Y.~Korolkov,  V.M.~Leontiev, V.P.~Novikov,
V.F.~Obraztsov,  V.A.~Polyakov, V.I.~Romanovsky, 
  V.I.~Shelikhov, N.E.~Smirnov,  M.M.~Soldatov, 
  O.G.~Tchikilev, E.~Usenko, M.V.~Vasiliev, E.V.~Vlasov, V.I.~Yakimchuk,
  O.P.~Yushchenko. }
  
\vskip 0.15cm
{\large\bf $Institute~for~High~Energy~Physics,~Protvino,~Russia$}

\vskip 0.35cm
{\bf V.N.~Bolotov, S.V.~Laptev, V.A.~Lebedev,  A.R.~Pastsjak, A.Yu.~Polyarush,  
  R.Kh.~Sirodeev.}
\vskip 0.15cm
{\large\bf $Institute~for~Nuclear~Research,~Moscow,~Russia$}
\vskip 0.15cm
\end{center}
\end{titlepage}
\newpage
\begin{center}
Abstract
\end{center}
 The  decay $K^{-} \rightarrow \mu^{-} \bar{\nu} \pi^{0} $ has been 
 studied using in-flight decays detected with "ISTRA+" setup operating
 in the 25 GeV negative secondary beam of the U-70 PS. About 112K events were
 used for the analysis. The $\lambda_{+}$  and  $\lambda_{0}$ slope
 parameters of the decay formfactors  $f_{+}(t)$, $f_{0}(t)$
  have been measured : \\
 $\lambda_{+}= 0.0321 \pm 0.004$(stat) $\pm 0.002$(syst) \\ 
 $\lambda_{0}= 0.0209 \pm 0.004$(stat) $\pm 0.002$(syst); 
 the correlation $d \lambda_{0}/d \lambda_{+}=-0.46$ \\ 
 The limits on the
 possible tensor and scalar couplings have been derived: \\
 $f_{T}/f_{+}(0)=-0.021 \pm 0.028$(stat) $\pm 0.014$(theory);  \\
 $f_{S}/f_{+}(0)=0.004 \pm 0.005$(stat) $\pm 0.005$(theory)
\newpage
\thispagestyle{empty}
\newpage
\raggedbottom
\sloppy

\section{ Introduction}
 The decay $K \rightarrow \mu \nu \pi^{0}$($K_{\mu 3}$) is known to be a key one 
 in hunting for phenomena beyond the Standard Model (SM).
 In  particular, significant efforts have been invested into T-violation searches
 by the measurements of the muon transverse polarization $\sigma_{T}$.
 In our analysis, based on $\sim$~112K events of the decay, we present new search for 
 S and T interactions  by fitting the $K_{\mu 3}$ Dalitz plot
 distribution, similar to as it was done for the $K_{e3}$ decay:
 \cite {paper1}.
    Another subject of our study is the measurement of the V-A
 $f_{+}(t)$ $f_{0}(t)$ formfactor slopes $\lambda_{+}$ and $\lambda_{0}$.
  
\section{ Experimental setup}
The experiment is performed at the IHEP 70 GeV proton synchrotron U-70.
The experimental setup  "ISTRA+" has been described in some details in our
recent paper on $K_{e3}$ decay \cite{paper1}. 
A schematic view of the detector is shown in Fig.1. 
 The setup is located in the 4A negative unseparated secondary beam. 
The beam momentum  is $\sim 25$ GeV with 
$\Delta p/p \sim 2 \%$. The admixture of $K^{-}$ in the beam is $\sim 3 \%$.
The beam intensity is $\sim 3 \cdot 10^{6}$ per 1.9 sec. U-70 spill.   
\begin{figure}
\epsfig{file=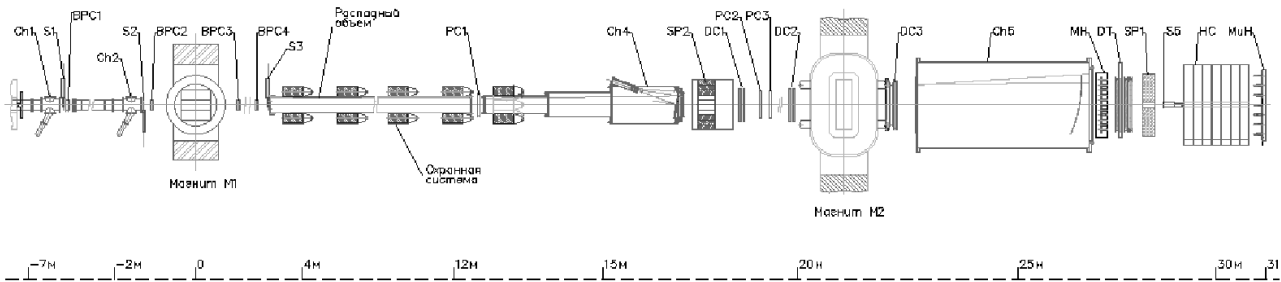,width=17cm}
\caption{ The layout of the   "ISTRA+" setup.}
\end{figure}
\section{Event selection}
During 3 weeks physics run in  March-April 2001, when the muon identification was in full
operation, 363M events were logged on DLT's. This information is supported
by about 100M MC events generated with Geant3 \cite{geant}. 
Some information on the  reconstruction procedure is presented in \cite{paper1},
here we touch only points relevant for the $K_{\mu 3}$ events selection.

The muon identification is based on the information from the 
$SP_{1}$ - a 576-cell lead glass calorimeter and 
 HC- a scintillator-iron sampling hadron calorimeter, subdivided
into 7 longitudinal sections 7$\times$7 cells each \cite{HCAL}. The calorimeters 
are located at the very end of the setup, after the main magnet (M2) 
and the last elements of the tracking system: drift tubes (DT) and the matrix
scintillation hodoscope (MH). The first requirement is that the energy of the 
$SP_{1}$ cluster, associated
with the charged track is less than $\sim$ 2.5 MIP's; the HC energy, associated with
the track should also be less than 2.5 MIP's. The last selection requires that more 
than
10$\%$ of the HC associated energy is deposited in 2 last layers (out of 7) of the HC.
The efficiency of the algorithm to muons is tested on $K \rightarrow \mu \nu$ events
and is found to be $\sim 70 \%$. The $\pi \rightarrow \mu $  misindification is
measured on $ K^{-} \rightarrow \pi^{-} \pi^{0}$ decay and  is $\sim 3 \%$.
After the muon identification, the selection of the events with  two extra showers 
results in $M_{\gamma \gamma}$ spectrum shown in Fig.2.
\begin{figure}
\begin{center}
\epsfig{file=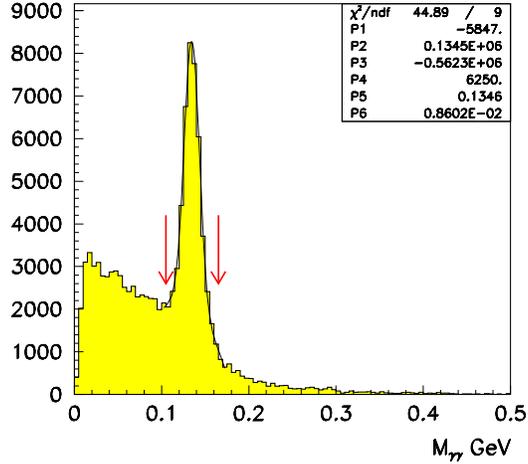,width=8cm}
\end{center}
\caption{ The $ \gamma \gamma$ mass spectrum for the events with
the identified muon and two extra showers.}
\end{figure}
The $\pi^{0}$ peak has a mass of $M_{\pi0}=134.6$ MeV, and a  resolution of 8.6 MeV.
The missing mass squared- $(P_{K}-P_{\mu}-P_{\pi^{0}})^{2}$,
where P are the corresponding four-momenta, is presented in Fig.3. The cut is 
$\pm 0.01$ GeV$^{2}$.
\begin{figure}
\begin{center}
\epsfig{file=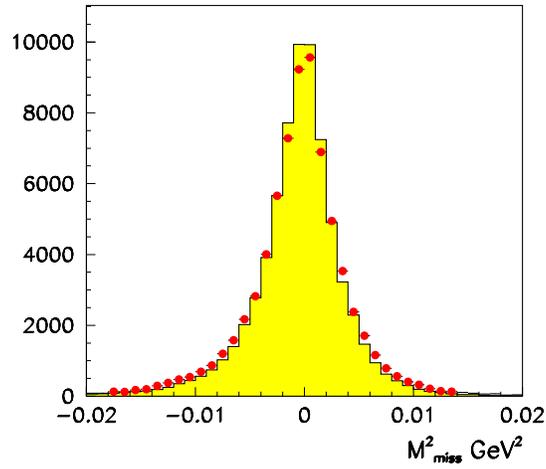,width=8cm}
\end{center}
\caption{The missing four-momentum squared $(P_{K}-P_{\mu}-P_{\pi^{0}})^{2}$
for the selected events . The points with errors are the data, the histogram- MC. }
\end{figure} 
The further selection  is done by the requirement that the event passes 2C   
 $K \rightarrow \mu \nu \pi^{0}$ fit.
The missing energy $E_{K}-E_{\mu}-E_{\pi^{0}}$ after this selection is shown 
in Fig.4
The peak at low $E_{miss}$ corresponds to
the remaining $K^{-} \rightarrow \pi^{-} \pi^{0}$ background. The corresponding
cut is $E_{miss}>1.4$ GeV.
 The surviving background is estimated from MC to be less
than $4 \%$.  
\begin{figure}
\begin{center}
\epsfig{file=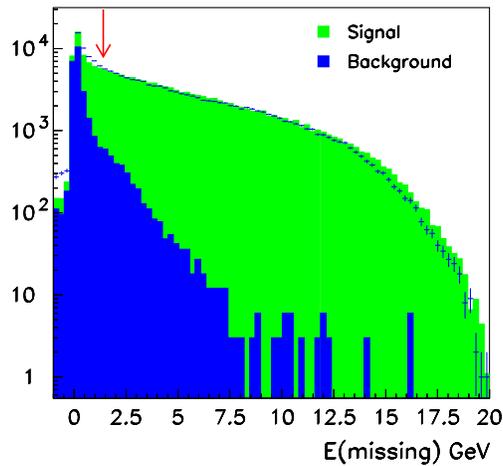,width=8cm}
\end{center}
\caption{The missing energy for the $\mu \pi^{0}$ events. 
 The points with errors are the data, the histograms- MC.
The dark(blue) peak at zero value corresponds to the 
MC-predicted $K \rightarrow \pi^{-} \pi^{0}$ background. 
The arrow indicates the cut value.}
\end{figure}
The detailed data reduction information is shown in Table.1.
\begin{table}[bth]
\caption{ Event reduction statistics. } 
\renewcommand{\arraystretch}{1.5}
\begin{center}
\begin{tabular}{|c|c|}
\hline
 Run  &           March-April     2001  \\  
\hline
 $N_{events}$ on tapes &  363.002.105 \\
\hline
Beam track reconstructed  &      268.564.958 =74 $\%$ \\
\hline
 One secondary track found &     134.227.095 =37$\%$  \\                    
\hline
 Written to DST  &   107.215.783 =30 $\%$ \\
\hline \hline
 $\mu^{-}$ identified and $\pi^{0}$ identified &      218.813 \\
\hline
 $|M_{miss}^{2}|<0.01$ &  195.799  \\
\hline

$K \rightarrow \mu \nu \pi^{0}$ accepted &  166.495 \\
\hline
$E_{miss} >$ 1.4 GeV &  112.157 \\
\hline
\end{tabular}
\end{center}
\end{table}
\renewcommand{\arraystretch}{1.0}

\section{ Analysis}
The event selection described in the previous section results in selected
112K events in  2001 data. The distribution of
the events over the Dalitz plot is shown in Fig.5. The  variables 
$y=2E_{\mu}/M_{K}$ and $z=2E_{\pi}/M_{K}$, where $E_{\mu}$, $E_{\pi}$ are the
energies of the muon and $\pi^{0}$ in the kaon c.m.s are used. 
\begin{figure}
\begin{center}
\epsfig{file=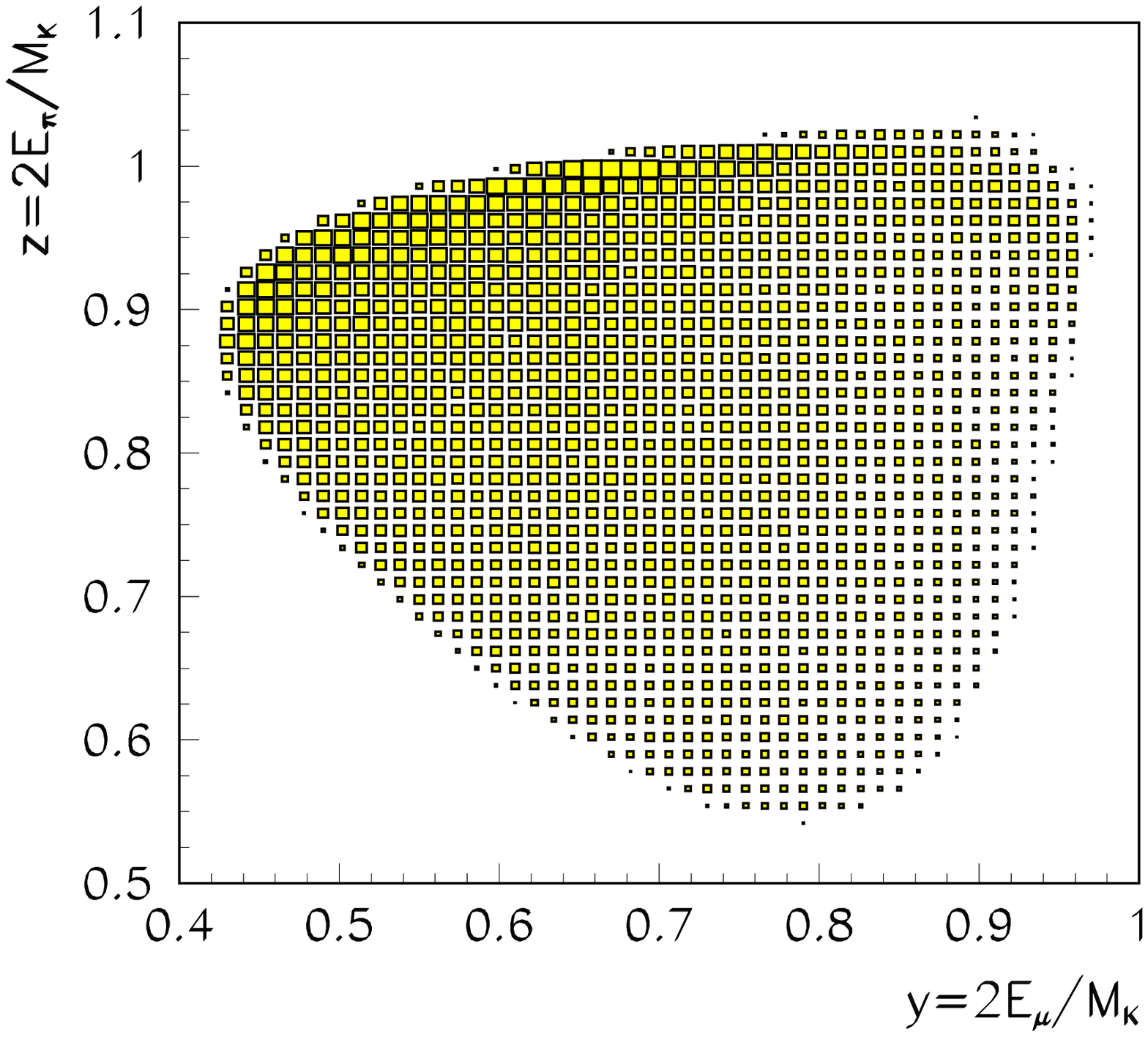,width=8cm}
\end{center}
\caption{ Dalitz plot $(y=2E_{\mu}/M_{K} ;z=2E_{\pi^{0}}/M_{K})$ for
the selected $K \rightarrow \mu \nu \pi^{0}$ events after the 2-C fit.}
\end{figure}
The most general Lorentz invariant form of the matrix element for the 
decay $K \rightarrow \mu \nu \pi^{0}$ is  \cite{Steiner}:
\begin{equation}
M= \frac{G_{F}sin\theta_{C}}{\sqrt{2}} \bar u(p_{\nu}) (1+ \gamma^{5})
[m_{K}f_{S} -
\frac{1}{2}[(P_{K}+P_{\pi})_{\alpha}f_{+}+
(P_{K}-P_{\pi})_{\alpha}f_{-}]\gamma^{\alpha} + i \frac{f_{T}}{m_{K}}
\sigma_{\alpha \beta}P^{\alpha}_{K}P^{\beta}_{\pi}]v(p_{\mu})
\end{equation}
It consists of scalar, vector and tensor terms. $f_{S}$, $f_{T}$, $f_{\pm}$
are  functions of $t= (P_{K}-P_{\pi})^{2}$. In the Standard Model (SM)
the W-boson exchange leads to the pure vector term. The "induced" 
scalar and/or tensor terms, due to EW radiative corrections are negligibly
small, i.e the nonzero scalar/tensor form factors indicate a physics
beyond SM. 

The term in the vector part, proportional to $f_{-}$ is reduced(using the Dirac
equation) to a scalar formfactor. In the same way, the tensor term is reduced to
a mixture of a scalar and a vector formfactors. The redefined $f_{+}$(V), 
$F_{S}$(S) and the corresponding Dalitz plot
density in the kaon rest frame $\rho(E_{\pi},E_{\mu})$ are \cite{Chizov}:
\begin{eqnarray}
 V & = & f_{+}+(m_{\mu}/m_{K})f_{T} \nonumber \\ 
S & = & f_{S} +(m_{\mu}/2m_{K})f_{-}+
\left( 1+\frac{m_{\mu}^{2}}{2m_{K}^{2}}-\frac{2E_{\mu}}{m_{K}}
-\frac{E_{\pi}}{m_{K}}\right) f_{T} \nonumber \\ 
\rho (E_{\pi},E_{\mu}) & \sim & A \cdot |V|^{2}+B \cdot Re(V^{*}S)+C \cdot |S|^{2} \\
A & = & m_{K}(2E_{\mu}E_{\nu}-m_{K} \Delta E_{\pi})-  
m_{\mu}^{2}(E_{\nu}-\frac{1}{4} \Delta E_{\pi}) \nonumber \\
B & = & m_{\mu}m_{K}(2E_{\nu}-\Delta E_{\pi}) \nonumber \\
C & = & m_{K}^{2} \Delta E_{\pi};~ \Delta E_{\pi}  =  E_{\pi}^{max}-E_{\pi} ;~
E_{\pi}^{max}= \frac{m_{K}^{2}-m_{\mu}^{2}+m_{\pi}^{2}}{2m_{K}} \nonumber 
\end{eqnarray}
Following \cite{Leutwyler} a scalar formfactor $f_{0}$ is introduced:
$f_{0}(t)=f_{+}(t)+ \frac{t}{m_{K}^{2}-m_{\pi}^{2}}f_{-}(t)$ and linear
dependence of $f_{+}, f_{0}$ on t is assumed:
 $f_{+}(t)=f_{+}(0)(1+\lambda_{+}t/m_{\pi}^{2})$;
 $f_{0}(t)=f_{+}(0)(1+\lambda_{0}t/m_{\pi}^{2})$. 
 Then $ f_{-}=f_{+}(0)(\lambda_{0}-\lambda_{+})
  \frac{m_{K}^{2}-m_{\pi}^{2}}{m_{\pi}^{2}}$.
 
The procedure for the experimental extraction of the parameters
$ \lambda_{+}$, $ \lambda_{0}$, $f_{S}$, $f_{T}$ starts from the subtraction of the 
MC estimated background from the Dalitz plot of Fig.4.
 The background 
normalization was determined by the ratio of the real and generated
$K^{-} \rightarrow \pi^{-} \pi^{0}$ events. Then the Dalitz plot was
subdivided into 20 $\times$ 20 cells.
The background subtracted  distribution of the numbers of  events in the cells
(i,j) over the Dalitz plot, for example, in the case of 
simultaneous extraction of $\lambda_{+}$, $\lambda_{0}$ and $\frac{f_{S}}{f_{+}(0)}$,
was fitted with the function:
\begin{eqnarray}
\rho (i,j)\sim \sum_{k_i;k_1+k_2+k_3=0,1,2 }W_{k_1 k_2 k_3}(i,j)
\cdot \lambda_{+}^{k_1}
\cdot \lambda_{0}^{k_2}
\cdot \left( f_{S}/f_{+}(0)\right)^{k_3}     
\end{eqnarray} 
Here $W_{k_1 k_2 k_3}$ are MC-generated  functions, which are build up as follows:
the MC events are generated with constant density over the Dalitz plot and
reconstructed with the same program as for the real events.  Each event
carries the weight w determined by the corresponding term in formula 2,  
calculated using the MC-generated("true") values for y and z.  
The radiative corrections according to \cite{grinb} were taken into account.
Then $W_{k_1 k_2 k_3}$ is constructed by summing up the weights w of the events in
the corresponding Dalitz plot cell. This procedure allows to avoid the
systematic errors due to the "migration" of the events over the Dalitz plot
because of the finite experimental resolution.
\section{Results}
The results of the fit are summarized in Table.2.
\renewcommand{\arraystretch}{1.5}

\begin{table}[bth]
\caption{ Results of the fit.}                                 
\begin{center}
\begin{tabular}{|c|cc|}
\hline
    & { $\mu^-\bar{\nu}\pi^0$}  &  
      { $\mu^-\bar{\nu}\pi^0 \; + \; e^-\bar{\nu}\pi^0$} \\ \hline\hline
    
  { $\lambda_+$} &  $~~0.0321^{+0.0040}_{-0.0040}$ & 
                         $~~0.0296^{+0.0014}_{-0.0014}$   \\ 
  { $\lambda_0$} &  $~~0.0197^{+0.0046}_{-0.0047}$ & 
                         $~~0.0209^{+0.0042}_{-0.0042}$   \\ \hline

  { $\lambda_+$} &  $~~0.0321^{+0.0040}_{-0.0040}$ & 
                         $~~0.0297^{+0.0014}_{-0.0014}$   \\    
  { $\lambda_0$} &  $0.01700$ & 
                        $0.01700$   \\ 
  { $ f_S/f_+(0)$}       &  $~~0.0034^{+0.0058}_{-0.0058}$ &
                         $~~0.0039^{+0.0052}_{-0.0052}$   \\ \hline
  		 
  { $\lambda_+$} &  $~~0.0338^{+0.0037}_{-0.0037}$ & 
                         $~~0.0299^{+0.0014}_{-0.0014}$   \\    
  { $\lambda_0$} &  $0.01700$ & 
                         $0.01700$   \\ 
  { $f_T/f_+(0)$}       &  $-0.0240^{+0.0330}_{-0.0326}$ &
                         $-0.0210^{+0.0278}_{-0.0274}$  \\ \hline\hline
			 
  { $\chi^2/\mbox{ndf}$} &  1.5  &  1.5 \\  \hline
  { $N_{\mbox{bins}}$} & 275    &  \\ \hline\hline
\end{tabular}
\end{center}

\end{table}

\renewcommand{\arraystretch}{1.0}

The first line corresponds to  pure V-A SM fit. The first column is independent
fit of our $K_{\mu 3}$ data. The $\lambda_{+} \div \lambda_{0}$ correlation parameter
is: $\frac{d\lambda_{0}}{d\lambda_{+}}=-0.46$.
The $\lambda_{+}$ value 
$\lambda_{+}^{\mu}=0.0321 \pm 0.004$
is in a good agreement with that, extracted from the analysis of our $K_{e3}$ data
\cite{paper1}:
$\lambda_{+}^{e}=0.0293 \pm 0.0015$, i.e our data do not contradict the $\mu -e$ 
universality. \\
In the second column the results of the joined fit of our
$K_{e3}$ and $K_{\mu 3}$ data
are presented(this is practically equivalent to fixing the $\lambda_{+}$ to it's
$K_{e3}$ value). This fit, of course, assumes the $\mu -e$ universality. The  $\lambda_{0}$
value $\lambda_{0}= 0.0209 \pm 0.0042$ is in a good agreement with the calculations
in the framework of the chiral perturbation theory($\chi PT$)\cite{Leutwyler}: 
$\lambda_{0}^{th}= 0.017 \pm 0.004$.\\
All the errors presented are from the "MINOS" procedure of the "MINUIT"
program \cite{Minuit} and are larger than the Gaussian ones.
At present, we estimate
an additional systematics error in $\lambda_{+}$, $\lambda_{0}$ to be $\pm 0.002$. 
The estimate is done by varying cuts, cell size during the fit of the Dalitz plots etc.

In the second and in the third lines the scalar and  the  tensor  terms are  added 
into the fit.  As it is seen from the second line of formula (2), the $f_{S}$ term is
$100 \%$ anti-correlated with the V-A contribution $(m_{\mu}/2m_{K})f_{-}$,
i.e an independent estimate of $f_{-}$ is necessary to extract $f_{S}$. By definition,
 $ f_{-}=f_{+}(0)(\lambda_{0}-\lambda_{+})\frac{m_{K}^{2}-m_{\pi}^{2}}{m_{\pi}^{2}}$.
 $\lambda_{+}$ is, in fact, defined by the $K_{e3}$ data, and  $\lambda_{0}$  
is calculated by $\chi PT$: $\lambda_{0}^{th}= 0.017 \pm 0.004$. In our $f_{S}$ fit
we fix $\lambda_{0}$ to this, theoretical, value. The error($\pm 0.004$)
in the theoretical prediction induces an additional error in $f_{S}$ equal to  
$\pm 0.005$. 

A possible example of  theories, which give nonzero $f_S$ are  the 2HDM \cite{2HDM} and
the Weinberg 3HDM model \cite{Weinberg}. In these theories,  $f_S$
comes from the diagram with the charged Higgs boson exchange $H^{-}$. The calculation
of the contributions gives \cite{Belanger}:
\begin{eqnarray}
f_{S}^{2hdm}/f_{+}(0)= \frac{m_{\mu}}{2m_{K}} \cdot \frac{m_{K}^{2}}{m_{H}^{2}}
\cdot tg^{2}(\beta) \\
f_{S}^{3hdm}/f_{+}(0)= \frac{m_{\mu}}{2m_{K}} \cdot \frac{m_{K}^{2}}{m_{H_{1}}^{2}}  
\cdot Re(\alpha_{1}^{*}\gamma_{1}) 
\end{eqnarray} 
Here $m_{H}$ is the charged Higgs-boson mass (mass of the lightest $H^{\pm}$
in case of 3HDM); $tg(\beta)=v_{2}/v_{1}$- the ratio of the vacuum expectation values
for 2 Higgs doublets;
$\alpha$ and $\gamma$ are complex couplings of the 3HDM Higgs boson to d-quarks and
leptons.\\ 
From our  limit for $f_{S}$ :
\begin{center} 
 $\frac{tg(\beta)}{m_{H}}= 0.39
  \pm 0.2(stat) \pm 0.2(theory) GeV^{-1}; $ \\[3mm]  
 $Re(\alpha_{1}^{*}\gamma_{1})\frac{m_{K}^{2}}{m_{H_{1}}^{2}}= 0.036
  \pm 0.047(stat) \pm 0.047(theory) $ 
\end{center}
 Our 2HDM limit is comparable with that from LEP searches for the decay
 $b \rightarrow \tau \nu_{\tau}$ \cite{LEP}:  $90 \%$ C.L. limit is
 $\frac{tg(\beta)}{m_{H}}< 0.4 \div 1$ GeV$^{-1}$(depending on collaboration).\\
 
 The  results of the fit with the tensor term are presented in the third line.
 The tensor term is also correlated with  $\lambda_0$, 
 $df_{T}/d \lambda_0=-3.5$. That's why we decided to apply the same approach
 for the tensor term as for the scalar one, i.e $\lambda_0$ is fixed to it's
 theoretical value and the induced error in $f_T$, due to the theoretical
 error in $\lambda_0$ is calculated. The error equals $ \pm 0.02$ for the
 single $K_{\mu 3}$ fit and  $ \pm 0.014$ for the combined one. \\
 The tensor coupling $f_{T}$ appears naturally in the leptoquark models, as a
 result of the Fierz transformation \cite{Belanger}. Unfortunately, we have not
 found complete theoretical consideration for this contribution.
\section{Summary and conclusions}
The $K^{-}_{\mu 3}$ decay has been studied using in-flight decays of 25 GeV 
$K^{-}$, detected by "ISTRA+" magnetic spectrometer. Due to the high
statistics, adequate resolution of the detector and good sensitivity over
all the Dalitz plot space, the measurement errors are significantly reduced
as compared with the previous measurements. 
 The $\lambda_{+}^{\mu}$ parameter of the vector formfactor $f_{+}(t)$
 is measured to be: 
\begin{center} 
 $\lambda_{+}^{\mu}= 0.0321 \pm 0.004(stat) \pm 0.002(syst)$.
\end{center}
and is in agreement with that obtained from our $K^{-}_{e 3}$ data:
\begin{center} 
 $\lambda_{+}^{e}= 0.0293 \pm 0.0015(stat) \pm 0.002(syst)$.
\end{center} 
The combined fit of both sets of data assuming the $\mu - e$ universality gives:
\begin{center} 
 $\lambda_{+}= 0.0296 \pm 0.0014(stat) \pm 0.002(syst)$.
\end{center} 
The $\lambda_{0}$ parameter of the scalar formfactor $f_{0}(t)$
 is measured to be:
 \begin{center} 
 $\lambda_{0}= 0.0209 \pm 0.004(stat) \pm 0.002(syst)$.
\end{center}
It is, at present, the best measurement of this parameter. It is in a good
agreement with $\chi PT$ prediction. 

   The limits on the
 possible scalar and tensor couplings are derived: 
\begin{center} 
 $f_{S}/f_{+}(0)=0.0039 \pm 0.005$(stat) $\pm 0.005$(theory) ;  \\[3mm]  
 $f_{T}/f_{+}(0)=-0.021 \pm 0.028$(stat) $\pm 0.014$(theory)   
\end{center}
The second(theoretical) error  comes from the 
uncertainty in the $\chi PT$ prediction for $\lambda_{0}$. Again, this is the
current best estimates for these parameters.

We would like to thank V.V.~Braguta, A.E.~Chalov, A.A.~Likhoded, A.K.~Likhoded, 
R.N.~Rogaliov, S.R.~Slabospitsky  for the discussions. 

The INR part of the collaboration 
is  supported by the RFFI fund contract N00-02-16074.

\end{document}